\documentclass[%
 reprint,
 showpacs,showkeys,
 amsmath,amssymb,
 aps,
]{revtex4-1}
\usepackage{amsmath}
\usepackage{cases}
\usepackage{amssymb}
\usepackage{CJK}
\usepackage{graphicx}
\usepackage{dcolumn}
\usepackage{bm}
\usepackage{color}
\usepackage[pagewise]{lineno}

\begin{document}

\preprint{APS/123-QED}

\title{Residue Coulomb interaction among isobars and its influence in symmetry energy of neutron-rich fragment}

\author{C. W. MA$^{1}$}\thanks{Email: machunwang@126.com}
\author{S. S. WANG$^{1,2}$}
\author{Y. L. ZHANG$^{1}$}
\author{H. L. WEI$^{1}$}

\affiliation{Institute of Particle and Nuclear Physics, Henan Normal University, \textit{Xinxiang 453007}, China\\
$^{2}$ Shanghai Institute of Applied Physics, Chinese Academy of Sciences, \textit{Shanghai 201800}, China
}

\date{\today}

\begin{abstract}
The residue Coulomb interaction (RCI), which affects the results of
symmetry energy of neutron-rich nucleus in isobaric yield ratio (IYR) methods, is
difficult to be determined. Four RCI approximations are investigated: (1) the M1--RCI
adopting the $a_{c}/T$ (the ratio of Coulomb energy coefficient to temperature)
determined from the IYR of mirror-nuclei fragment; (2) the M2--RCI by fitting the
difference between IYRs; (3) the M3--RCI by adopting the standard Coulomb energy at a
temperature $T=2$MeV; and (4) neglecting the RCI among the three isobars. The M1--, M2-- and M3--RCI is found to no larger than
0.4. In particular, the M2--RCI is very close to zero. The effects of RCI in the $a_{sym}/T$ of
fragment are also studied. The M1-- and M4--$a_{sym}/T$ are found to be the lower and
upper limitations of $a_{sym}/T$, respectively. The M2--$a_{sym}/T$ overlaps the M4--$a_{sym}/T$,
which indicates that the M2--RCI is negligible, at the same time the RCI among the three isobars can be neglected.
A relative consistent low values of M3--$a_{sym}/T$ ($7.5\pm2.5$) are found in very neutron-rich isobars.
\end{abstract}

\pacs{21.65.Cd, 25.70.Pq, 25.70.Mn}
\keywords{isobaric yield ratio, residue Coulomb interaction, symmetry energy, neutron-rich nucleus}
\maketitle


\section{introduction}
The study of nuclear symmetry energy (NSE) has continuously attracted attention
because of its importance in nuclear physics and astrophysics. Lots of theoretical
and experimental methods are proposed to investigate the NSE of nuclear matters
from sub-saturation to supra-saturation densities, which can both be produced in
heavy-ion collisions (HICs) \cite{BALi08PR}. Among the various observable to study
NSE --- such as flow \cite{Esymflows}, emission of light particles ($n, p$ and
the ratio between them) \cite{Kumar12PRCRnp,npDnpratio,DongJM13PRC}, isospin
diffusion \cite{Tsang-isodif-PRL04}, neutron-skin thickness \cite{skinSym}, isoscaling
(for colliding source or fragments) \cite{Isoscaling,Bot02-iso-T,MBTsPRL01iso,HShanPRL,Fang07-iso-JPG,FuY09isoCaNi,Iso-fluctuation13,ChenZQ10-iso-sym,OnoAMDRRCiso,SouzaPRC09isot,Soul03-iso,Soul06-iso-T-sym,TianWD05-iso-CPL,ZhouPei11T} or $m$-scaling \cite{Huang10NPA-Mscaling,PMar-IYR-sym13PRC,PMar12PRCIsob-sym-isos},
isobaric yield ratio (IYR) \cite{PMar12PRCIsob-sym-isos,Huang-PRC11-freeenergy,Huang10,MaCW11PRC06IYR,NST13Lin,NST13WADA},
etc. --- the IYR methods have attracted much attention recently \cite{PMar-IYR-sym13PRC,PMar12PRCIsob-sym-isos,Huang-PRC11-freeenergy,Huang10,MaCW11PRC06IYR,NST13Lin,NST13WADA,Mallik13-sym-IYR,MaCW12PRCT,MaCW13PRCT,MaCW12EPJA,MaCW12CPL06,MaCW13CPC,MaCW13isoSB,Ma13finite}. In IYR, parameters depending only on the mass of fragment cancel out, which makes the IYR methods possible to
study the symmetry energy of neutron-rich fragment specifically. In some similar isobaric
analysis, the symmetry energy of nucleus (including the volume symmetry energy and surface
symmetry energy) has also been investigated \cite{isobsymmNPA07,MeiHJPG10,MA12CPL09IYRAsbsbv}.

Different to the uniform value of parameter in mass formula, the symmetry-energy coefficient
($a_{sym}$) of nucleus or fragment in isobaric methods becomes nonuniform, and depends on the
neutron-excess ($I\equiv N-Z$) and mass number $A$ \cite{Huang10,MaCW12EPJA,MaCW12CPL06,MaCW13CPC,isobsymmNPA07,MeiHJPG10,MA12CPL09IYRAsbsbv}.
$a_{sym}$ is found to decrease with increasing $I$ in isobaric chains \cite{MA12CPL09IYRAsbsbv},
and similar phenomena is also shown in fragments produced in HICs \cite{Huang10,MaCW12EPJA,MaCW12CPL06}.
The difference between $a_{sym}$ (or $a_{sym}/T$) of isobars is found to decrease when the
nucleus (fragment) becomes more neutron-rich. In particular, $a_{sym}$ ($a_{sym}/T$) of nucleus (fragment) with
large neutron-excess is found to be very similar \cite{MaCW12EPJA,MA12CPL09IYRAsbsbv,Ma2013NST}.
It is also shown that the volume-symmetry-energy coefficient ($b_v$) and the surface-symmetry-energy
coefficient ($b_s$) both decrease with increasing $I$, but tend to be very similar in the very
neutron-rich nucleus (when $I > 13$) \cite{MA12CPL09IYRAsbsbv}.

In IYR methods to determine the symmetry energy of fragment, the residue Coulomb interaction
(RCI) between isobars is also difficult to be known due to the difficulty to separate the
temperature and energy terms contributing to free energy. In previous works, the standard
Coulomb term is used to calculate RCI \cite{isobsymmNPA07,MeiHJPG10,MA12CPL09IYRAsbsbv}, or the
value of $a_c/T$ (the ratio of Coulomb energy coefficient to temperature) for mirror nuclei
[IYR(m)] is used as an approximation for more neutron-rich fragments. For examples, the value
of $a_c/T$ is determined from the scaling between the IYR(m) and the $Z/A$ of the reaction
systems in a series of reactions, and $a_c/T$ is used for the $I=3$ fragments \cite{Huang10}.
Though $a_c/T$ can also be fitted from IYR(m) in a single reaction \cite{MaCW11PRC06IYR,MaCW12EPJA,MaCW12CPL06}.
it is supposed to be influenced by the volume or mass of projectile \cite{Souza12finite}.
But it is also suggested that the volume dependence of $a_c/T$ disappears
in the reactions induced by neutron-rich projectiles \cite{Ma13finite}. The different
RCI approximations request the comparison between them, and the study of the RCI effects
in the resultant symmetry energy of neutron-rich fragment is also required, which will
be focused on in this article. The article is organized as follows: Sec. \ref{model}
describes the IYR method and the approximations of the RCI; Sec. \ref{result} discusses
the results of the different RCIs and their effects in the symmetry energy of fragment
in measured reactions; Sec. \ref{summary} presents the summary of the article.

\section{model description}
\label{model}
In free energy based theories, the yield of fragment is determined by its
free energy, the property of colliding source, temperature, etc. in HICs above
the Fermi energy \cite{PMar12PRCIsob-sym-isos,AlbNCA85DRT,ModelFisher1,ModelFisher3}.
In a modified Fisher model (MFM), the free energy of a cluster (fragment) equals
its binding energy at nonzero temperature, which includes the contribution of
entropy \cite{ModelFisher3}. This makes the IYR method can conveniently determine
the symmetry energy of fragment. The description of IYR
methods can be partly found in Refs. \cite{PMar12PRCIsob-sym-isos,Huang-PRC11-freeenergy,Huang10,MaCW11PRC06IYR,MaCW12PRCT,MaCW12EPJA,MaCW12CPL06}.
For a better understanding of the methods, first we briefly describe the IYR methods
in MFM. Then the methods dealing with RCI will be intensively described.

In MFM the yield of a fragment with mass $A$ and neutron-excess $I$,
$Y(A,I)$, is given by \cite{ModelFisher1,ModelFisher3},
\begin{eqnarray}\label{Y}
Y(A,I) &= CA^{-\tau}exp\{[W(A,I)+\mu_{n}N+\mu_{p}Z]/T \nonumber\\
&+N\mbox{ln}(N/A)+Z\mbox{ln}(Z/A)\},
\end{eqnarray}
where $C$ is a constant; $A^{-\tau}$ originates from the entropy of fragment; $\tau$ is
independent of fragment size, but is nonuniform in different reactions \cite{ModelFisher3,Huang10Powerlaw}.
$\mu_n$ and $\mu_p$ are the neutron and proton chemical potentials, respectively; $T$ is
the temperature, and $W(A,I)$ is the free energy of cluster at $T$, which equals the binding
energy of the cluster. At a given $\rho$ and $T$, $W(A,I)$ can be parameterized as the
Weisz\"{a}cker-Bethe form mass formula  \cite{Weiz-Bethe},
\begin{eqnarray}\label{W}
W(A,I)&= a_v(\rho,T)A-a_s(\rho,T)A^{2/3}-E_{c}(\rho,T)   \nonumber\\
&-a_{sym}(\rho,T)I^2/A-\delta(N,Z),
\end{eqnarray}
where the indices $v, s,$ and $sym$ represent the volume-, surface-, and symmetry- energy,
respectively. $E_{c}(\rho,T)$ represents the Coulomb energy (assuming a spherical expansion,
at low densities the Coulomb energy decreases as $\rho^{1/3}$). The coefficients contain
contributions both from the binding energy and the entropy of the cluster due to nonzero
$T$ \cite{ModelFisher3}. For simplification, $a_i(\rho,T)$ is written as $a_i$ ($i$ represents
the different indices).

Inserting Eq. (\ref{W}) into Eq. (\ref{Y}), the IYR between isobars differing by 2 units in
$I$, $R(I+2,I,A)$, can be defined as,
\begin{eqnarray}\label{ratiodef}
&R(I+2,I,A)=Y(A,I+2)/Y(A,I) \hspace{1.5cm} \nonumber\\
&={exp}\{[W(I+2,A)-W(I,A)+(\mu_n-\mu_p)]/T \nonumber\\
&+S_{mix}(I+2,A)-S_{mix}(I,A)\},
\end{eqnarray}
where $S_{mix}(I,A) = N\mbox{ln}(N/A) + Z\mbox{ln}(Z/A)$. Assuming that $a_v, a_s$, $\mu_n$, and $\mu_p$
for the $I$ and $I + 2$ isobars are the same, inserting Eq.~(\ref{W}) into Eq.~(\ref{ratiodef}), and
taking logarithm of the resultant equation, one gets the IYR for odd-$I$ isobars,
\begin{eqnarray}\label{lnRcorio}
\mbox{ln}[R(I+2,I,A)] -\Delta_{I}&=[\Delta\mu-4a_{sym}(I+1)/A   \nonumber\\
&-\Delta E_{c}(I+2,I,A)]/T,
\end{eqnarray}
where $\Delta_{I} = S_{mix}(I + 2,A) - S_{mix}(I,A)$; $\Delta E_{c}(I+2,I,A) = E_{c}(I+2) - E_{c}(I)$;
$\Delta\mu = \mu_n - \mu_p$, $Z$ is the charge numbers of
the reference nucleus $(A,I)$. The pairing energy in Eq. (\ref{lnRcorio}) is
avoided in the odd-$I$ isobars. It is hard to know $\Delta\mu$, $a_{sym}$, $\Delta E_{c}$, and
$T$ exactly, because of the difficulty to separate them in the ratios. $\Delta\mu$ is assumed to change slowly
in the reactions, thus $\Delta\mu/T$ can be the same in $R(I+2,I,A)$ and $R(I,I-2,A)$.
Then $\Delta E_{c}/T$ is the retained term that affects $a_{sym}/T$ in the IYR methods.
Taking the difference between the neighboring IYRs,
\begin{eqnarray}\label{DifIYR}
&\mbox{ln}[R(I,I-2,A)]-\mbox{ln}[R(I+2,I,A)]-\Delta_{21} \nonumber\\
&=8a_{sym}/(AT)-\Delta E_{c21}/T,
\end{eqnarray}
where $\Delta_{21} = \Delta_{I-2}-\Delta_{I}$, and $\Delta E_{c21} = \Delta E_{c}(I,I-2,A) - \Delta E_{c}(I+2,I,A)$
is the RCI among the (I+2,A), (I, A) and (I-2,A) isobars. Since $\Delta E_{c21}/T$ can be viewed as one parameter, it is also
called as RCI between IYRs, and it is the parameter that will be focused on in this article. For neutron-rich
fragment, $a_{sym}/T$ can be obtained from Eq. (\ref{DifIYR}) as follows \cite{Huang10,MaCW12EPJA,MaCW12CPL06,MaCW13CPC},
\begin{eqnarray}\label{asym3}
a_{sym}/T&=\frac{A}{8}\{\mbox{ln}[R(I,I-2,A)]-\mbox{ln}[R(I+2,I,A)]   \nonumber\\
&-\Delta_{21}+\Delta E_{c21}/T\},
\end{eqnarray}

\begin{figure*}
\begin{center}
\includegraphics[width=12cm]{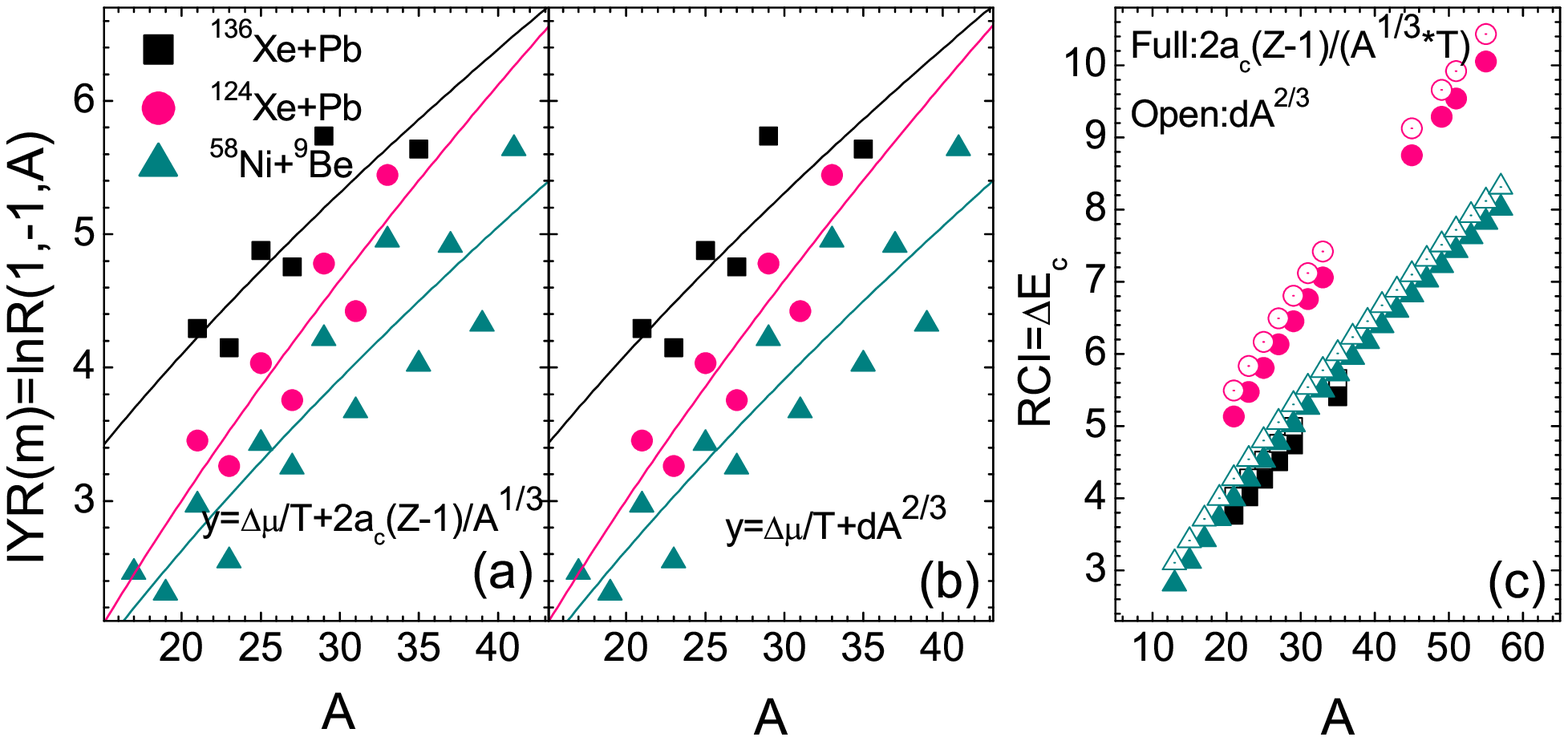}
\caption{(Color online) The isobaric yield ratio for mirror nuclei [IYR(m)]
in the 140$A$ MeV $^{58}$Ni + $^9$Be reactions \cite{Mocko06} and the 1$A$ GeV $^{124,136}$Xe + Pb
reactions \cite{Henz08}. In (a) the fitting function is according to Eq. (\ref{acmirror}), and in
(b) the fitting use the function using $y=\Delta\mu/T+dA^{2/3}$. (c) RCI between the mirror nuclei,
with full and open symbols representing RCIs according to panels (a) and (b).
}\label{MirrorFit}
\includegraphics[width=12cm]{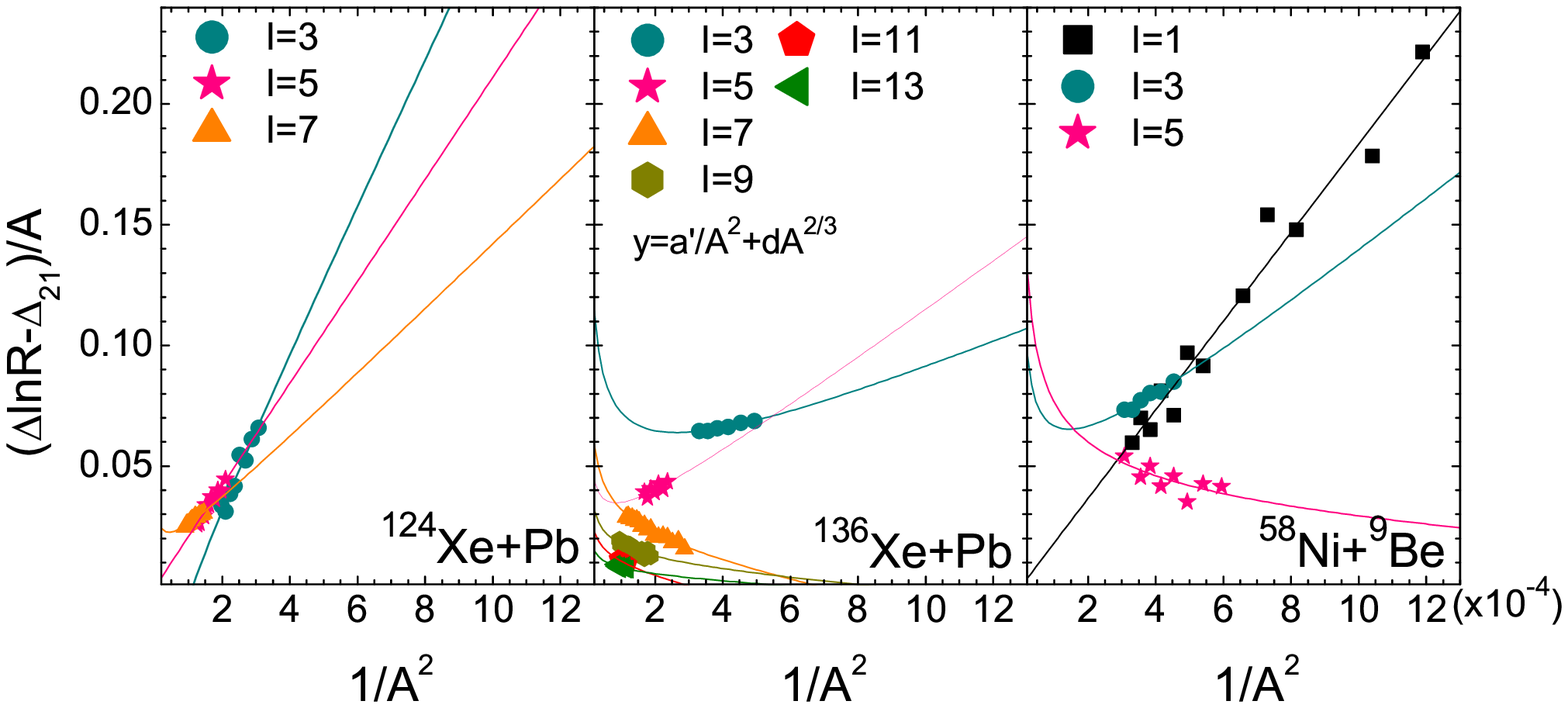}
\caption{(Color online)  The correlation between $\Delta\mbox{ln}R/A$
and $1/A^2$ of fragment in the 1$A$ GeV $^{124,136}$Xe + Pb \cite{Henz08},
and 140$A$ MeV $^{58}$Ni + $^9$Be reactions \cite{Mocko06}.
The lines represent the fitting results according to Eq. (\ref{adfit}).
}\label{dinXePb}
\end{center}
\end{figure*}

As mentioned above, due to the RCI ($\Delta E_{c}/T$ or through $a_{c}/T$) is hard to be known, four approximations
are used to deal with RCI in previous works,
\begin{description}
  \item[(i)] M1: $\Delta E_{c}/T$ (or $a_{c}/T$) is obtained from IYR(m). The following
       equation is used to extract the $\Delta E_{c}/T$ and $\Delta\mu/T$ based on Eq. (\ref{lnRcorio}),
       \begin{equation}\label{acmirror}
       \mbox{IYR(m)}=\mbox{ln}[R(1,-1,A)]=(\Delta\mu-\Delta E_{c})/T,
       \end{equation}
      The value of $\Delta\mu/T$, at the same time $\Delta E_{c}/T$ (or $a_{c}/T$)
      for the mirror nuclei can be determined. Assuming that $a_{c}/T$ are the same for all the
      fragments, $\Delta E_{c21}/T$ in Eq. (\ref{asym3}) is known \cite{MaCW11PRC06IYR,MaCW12EPJA,MaCW12CPL06,Ma13finite}.
  \item[(ii)] M2: $\Delta E_{c21}/T$ is determined from the difference between IYRs. Considering
        the free energy per particle at $T$ and pressure $P$ in MFM, the ratio of free energy
        to temperature near the critical point can be expanded as \cite{PMar12PRCIsob-sym-isos,Huang-PRC11-freeenergy},
        \begin{eqnarray}\label{FEpAtoT}
        \Psi(m_f,A,T,H)/T&=\frac{1}{2}am^{2}_{f}+\frac{1}{4}bm^{4}_{f}+\frac{1}{6}cm^{6}_{f} \nonumber\\
        &-\frac{H}{T}m_{f}+o(m^{8}_{f}),
        \end{eqnarray}
       where the parameters $a, b$ and $c$ depend on $T$ and $\rho$, and are used for fitting; $m_f = I/A$.
       $H$ is the conjugate field. The free energy is even in the exchange of $m_{f}\rightarrow-m_{f}$,
       reflecting the invariance of nuclear forces when exchanging $N$ and $Z$. This symmetry is violated
       by $H$, which arises when the source is asymmetric in chemical composition. $H$ and $m_{f}$ are
       related to each other through the relation $m_{f} = \frac{\delta F}{\delta H}$ \cite{Huang-PRC11-freeenergy}.
       The Coulomb energy for large $Z$ nucleus can be written as,
        \begin{equation}\label{Coulomb1}
        \frac{E_c}{A}=0.77\frac{Z^2}{A^2}A^{2/3}=\frac{0.77}{4}(1-m_{f})^{2}A^{2/3},
        \end{equation}
      Adding this term to the free energy $\Psi/T$, a quadratic and linear term in $m_{f}$ are introduced, which
      modify the symmetry energy coefficient and $H$. Also a term independent on $m_f$ is introduced \cite{Huang-PRC11-freeenergy}. The RCI
      is relevant only in the calculation of $a_{sym}/T$ for the large mass fragments in the IYR method,
      thus the $o(m^4)$ terms are negligible. At the same time, $(1/A^2)\varpropto m^2$. According to Eq. (\ref{ratiodef}), the fit of the
      quantity $[\mbox{ln}R(I,I-2,A)-\mbox{ln}R(I+2,I,A)]/A$ allows the estimation of the fitting parameters
      in Eq. (\ref{FEpAtoT}) and the Coulomb term. 
      The fitting function between the difference of IYRs and RCI is \cite{PMar12PRCIsob-sym-isos},
        \begin{equation}\label{adfit}
        \Delta\mbox{ln}R/A=a'/A^2+dA^{2/3},
        \end{equation}
      where $\Delta\mbox{ln}R = \mbox{ln}[R(I,I-2,A)]-\mbox{ln}[R(I+2,I,A)]$. $a'$ and $d$ are fitting
      parameters. The $a_{sym}/T$ of neutron-rich fragments can be obtained as,
      \begin{equation}\label{M2eq}
      \frac{a_{sym}}{T}=\frac{A}{8}(\Delta\mbox{ln}R-\Delta_{21})-dA^{2/3},
      \end{equation}
      The $dA^{2/3}$ term serves as the RCI. 

  \item[(iii)] M3: $\Delta E_{c21}$ is calculated by adopting
      $E_{C}=\frac{3}{5}\frac{Z^2e^2}{1.2A^{1/3}}[1-\frac{5}{4}(\frac{3}{2\pi})^{2/3}\frac{1}{Z^{2/3}}]$ \cite{MeiHJPG10,MA12CPL09IYRAsbsbv},
      and using $T=2$MeV since a relative low temperature is obtained using similar IYR methods \cite{MaCW12PRCT,MaCW13PRCT}. In this method, the RCI is
      \begin{equation}\label{M4RCI}
      \Delta E_{c21}/T=[E_c(I-2)+E_c(I+2)-2E_c(I)]/T,
      \end{equation}

  \item[(iv)] M4: We adopt a new approximation in this article, i.e., the RCI is omitted since it is the difference between three isobars, which is
      different to the RCI between two isobars and can be supposed to be negligible (the RCI between two isobars depends both on the $m_f^2$ and $m_f$).

\end{description}

\section{results and discussion}
\label{result}

The yields of fragments in the 140$A$ MeV $^{58}$Ni + $^9$Be reactions, which were measured
by Mocko \textit{et al.} at the National Superconducting Cyclotron Laboratory (NSCL)
in Michigan State University \cite{Mocko06}, and in the 1$A$ GeV $^{124,136}$Xe + Pb
reactions, which were measured by Henzlova \textit{et al.} at the FRagment Separator (FRS),
GSI Darmstadt \cite{Henz08}, will be used to perform the analysis.

First, we determine $a_{c}/T$ from IYR(m) for M1. In Fig. \ref{MirrorFit}(a), the IYR(m) is plotted
(similar results can be found in Refs. \cite{MaCW12EPJA,MaCW12CPL06,Ma13finite}). The form of
Coulomb energy adopted is $E_{c}=a_{c}Z(Z-1)/A^{1/3}$, and the resultant fitting function
being $y = \Delta\mu/T+2a_c(Z-1)/(A^{1/3}T)$ according to Eq. (\ref{acmirror}), in which the
quantity $2a_c(Z-1)/(A^{1/3}T)$ serves as RCI in IYR(m). In Fig. \ref{MirrorFit}(b), the fitting
function is modified to $y=\Delta\mu/T+dA^{2/3}$ due to $(Z-1)/A^{1/3}\propto A^{2/3}$,
and $a_c/T$ is assorted to the parameter $d$, in which $d\cdot A^{2/3}$ serve as RCI in
IYR(m). The resultant RCIs according to panels (a) and (b) are plotted in Fig. \ref{MirrorFit}(c), which
has a relative little difference of no larger than 0.2.

\begin{figure}
\begin{center}
\includegraphics[width=8cm]{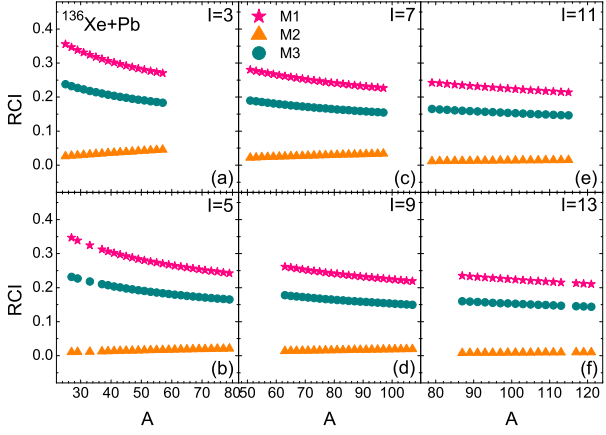}
\caption{(Color online)
The RCIs between isobars in the 1$A$ GeV $^{136}$Xe + Pb reaction using the M1--M3 approximations.
}\label{RCI3}
\end{center}
\end{figure}

Second, $d$ in M2 is determined by fitting the difference between IYRs according
to Eq. (\ref{adfit}). In Fig. \ref{dinXePb}, the correlation between $(\Delta\mbox{ln}R-\Delta_{21})/A$
and $1/A^2$ of fragments in the 1$A$ GeV $^{124,136}$Xe + Pb and 140$A$ MeV $^{58}$Ni + $^9$Be
reactions are plotted. The data can be well fitted by Eq. (\ref{adfit})
in the three reactions. But the distributions of the $(\Delta\mbox{ln}R-\Delta_{21})/A\sim 1/A^2$
correlation show different trend in fragments of the same $I$ in the three reactions, and the
same phenomena also happens in fragments of different $I$ in the same reaction.


Furthermore, taking the fragments in the 1$A$ GeV $^{136}$Xe + Pb reaction as an example, the RCI
in the M1--M3 approximations are compared, which are plotted in Fig. \ref{RCI3}. The values of RCI
show that M1 $>$ M3 $>$ M2, with M1-- and M3--RCI are no larger than 0.4, and the M2--RCI very close
to zero. The results verify that in Eq. (\ref{adfit}) the parameter $a'$ includes almost the whole
real RCI relating to the $m_{f}^2$ term, which results in M2--RCI is very small and can
be neglected.

\begin{figure}
\begin{center}
\includegraphics[width=8.6cm]{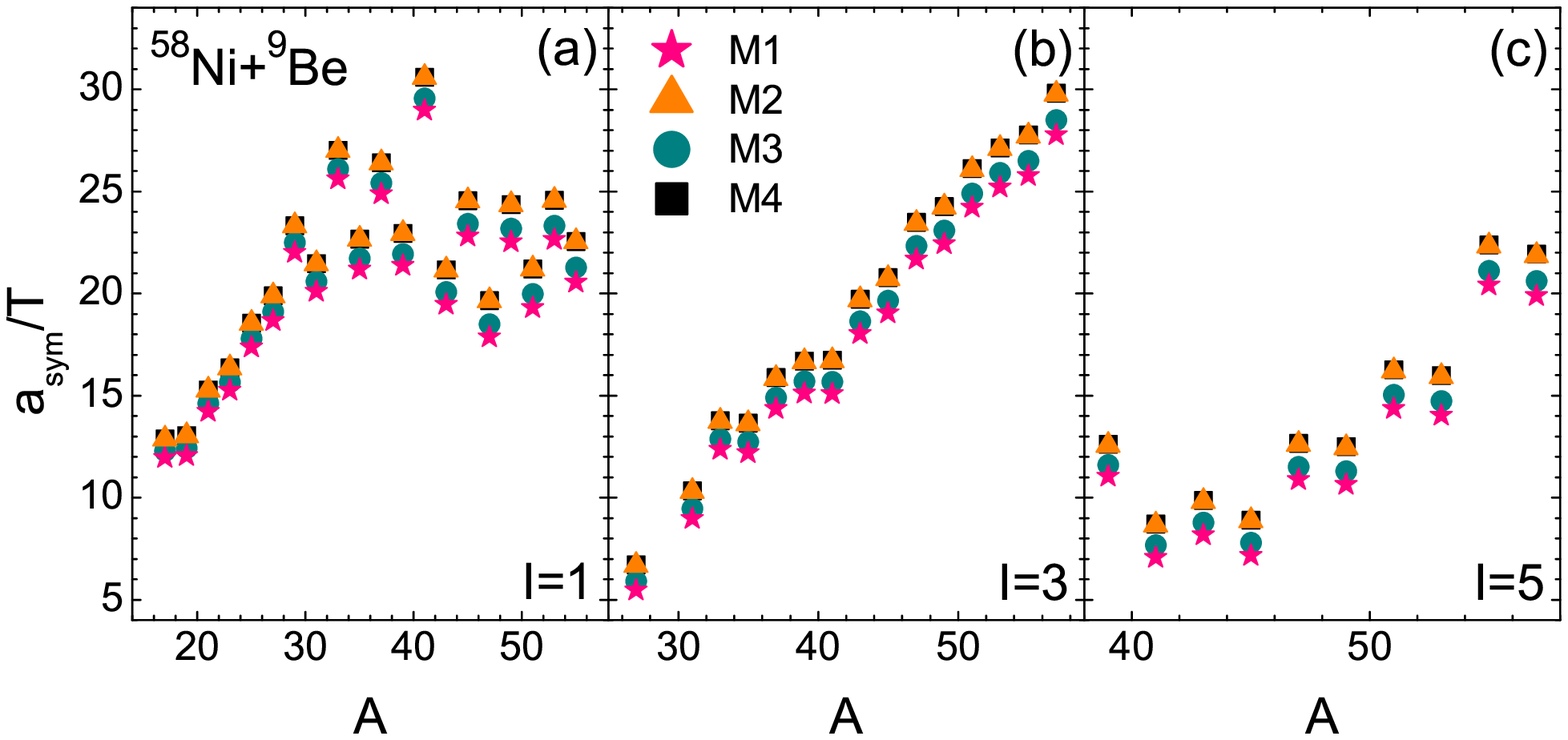}
\caption{(Color online) The $a_{sym}/T$ of fragments in the 140$A$ MeV
$^{58}$Ni + $^{9}$Be reactions. The labels M1 (star), M2 (triangle), M3 (circle), and M4 (square) denote the
$a_{sym}/T$ of fragments using the corresponding RCI.}
\label{CompasymTNi}
\includegraphics[width=8.6cm]{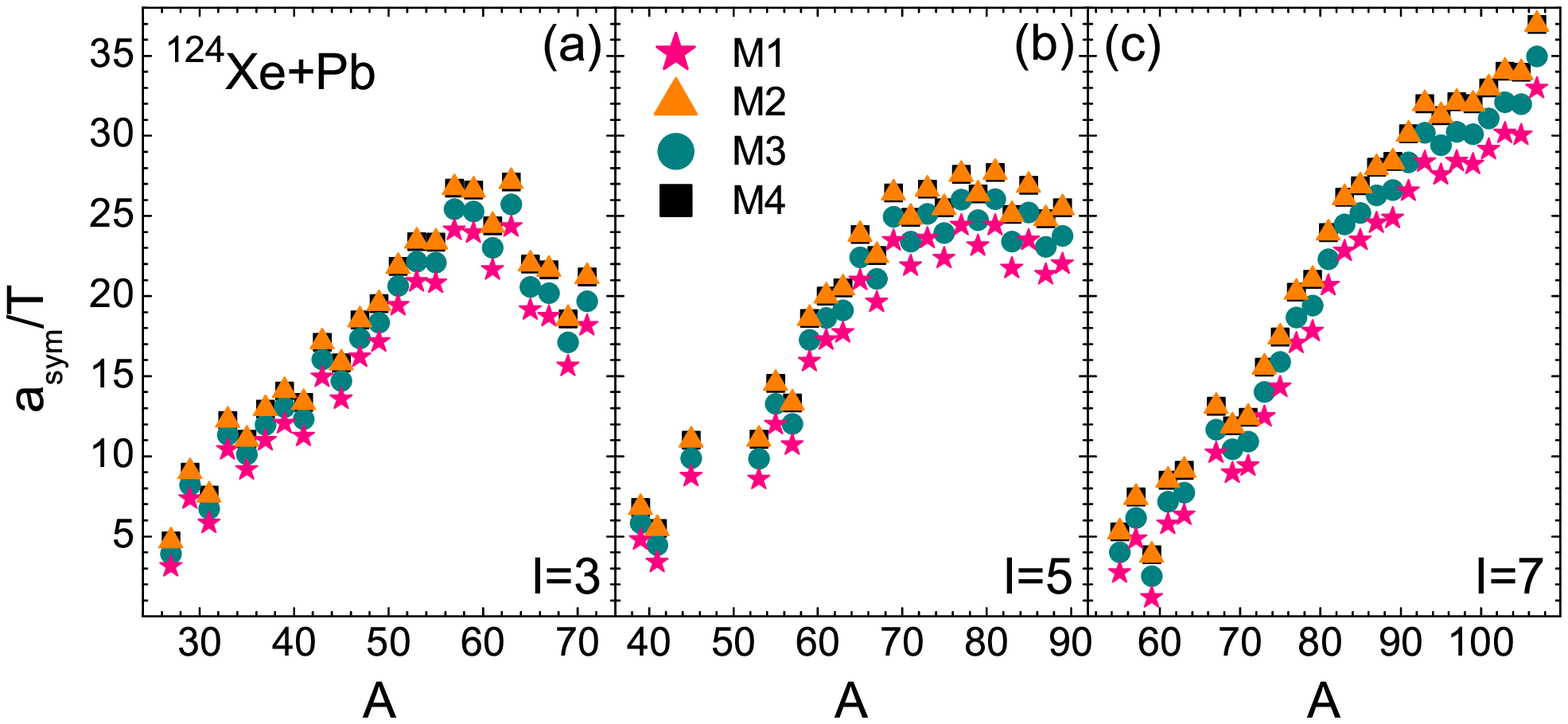}
\caption{(Color online) The same as in Fig. \ref{CompasymTNi} but
for the fragments in the 1$A$ GeV $^{124}$Xe + Pb reactions.
}\label{CompasymTXe124}
\includegraphics[width=8.6cm]{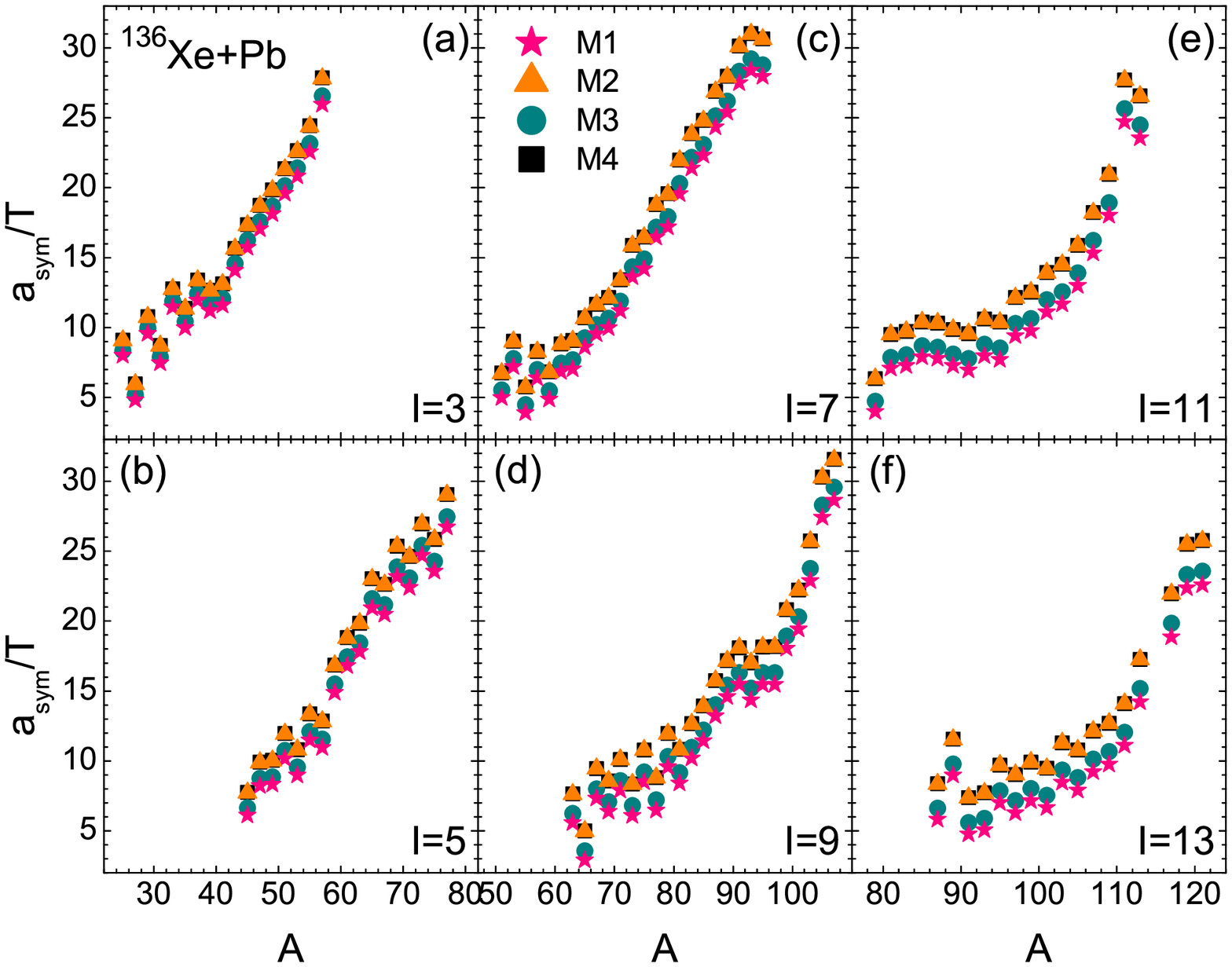}
\caption{(Color online) The same as in Fig. \ref{CompasymTNi} but
for the fragments in the 1$A$ GeV $^{136}$Xe + Pb reactions.
}\label{CompasymTXe136}
\end{center}
\end{figure}

It is important to study the effects of RCI in the resultant $a_{sym}/T$ of neutron-rich fragments.
The M1-- and M3--$a_{sym}/T$ are calculated according to Eq. (\ref{asym3}) using the corresponding
RCI, and the M2--$a_{sym}/T$ according to Eq. (\ref{M2eq}). First, the $a_{sym}/T$ of fragments in
the $^{58}$Ni reaction are plotted in Fig. \ref{CompasymTNi} and those of fragments in the $^{124, 136}$Xe
reactions are plotted in Fig. \ref{CompasymTXe124} and Fig. \ref{CompasymTXe136}, respectively.
$a_c/T =$0.55, 0.71 and 0.52 are used for M1 for the $^{58}$Ni, $^{124}$Xe and $^{136}$Xe reactions
is used, respectively. Generally, the M1--$a_{sym}/T$ and M4--$a_{sym}/T$ form the lower and upper
limitations of $a_{sym}/T$. The M1--$a_{sym}/T$ being the lower limitation of $a_{sym}/T$
indicates that the direct use of $a_c/T$ from IYR(m) may underestimate the real value of $a_{sym}/T$
in some degree. The M2--$a_{sym}/T$ almost overlaps the M4--$a_{sym}/T$. Since the M2--$a_{sym}/T$
incorporates the $m_f^2$ term of RCI (which is called as the effective symmetry energy \cite{Huang-PRC11-freeenergy}),
the result also indicates that the $m_f^2$ dependence of RCI is also very small, and the M2-- and
M4--$a_{sym}/T$ are close to the actual value. Thus the omission of RCI among three isobars is
reasonable in determining the coefficient of symmetry energy of neutron-rich fragment using
Eq. (\ref{asym3}).

\begin{figure}
\begin{center}
\includegraphics[width=8cm]{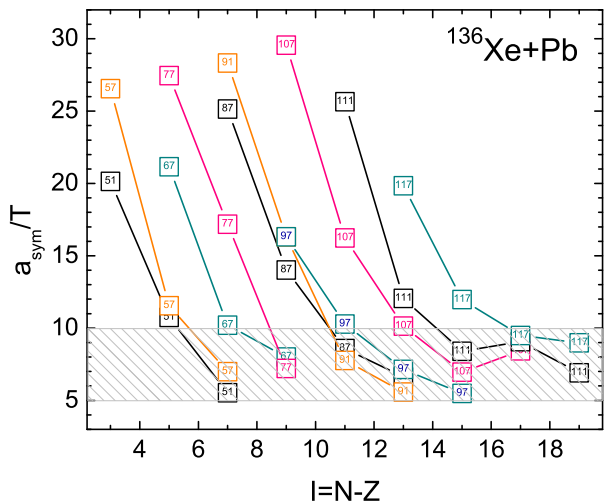}
\caption{(Color online) The values of $a_{sym}/T$ for fragments using the M3--RCI
(omitting the RCI) in the 1$A$ GeV $^{136}$Xe + Pb reactions. The number in the square denotes the
mass number of the fragment.
}
\label{Xe-Asym-T}
\end{center}
\end{figure}

To see the evolution of $a_{sym}/T$ in isobars, for a series of isobars in the 1$A$ GeV
$^{136}$Xe + Pb reactions, the values of M3--$a_{sym}/T$ are re-plotted in Fig. \ref{Xe-Asym-T}
since the M3--$a_{sym}/T$ is not influenced by the fitting parameters of RCI or the size of the
reaction system. The re-plotted isobars are from $A =$ 57 to 107 in the step of 10, plus
the $A =$ 51, 91, and 111 isobars. For each isobaric chain, $a_{sym}/T$ decreases as the
fragment becomes more neutron-rich. But a relative consistent values of $7.5\pm2.5$ is found
in the neutron-rich fragments for all the plotted isobaric chains, which are shown as the
shadowed area.

We will comment on the question raised in the result of $A$ and $I$ dependence of $a_{sym}/T$.
Actually, in isobaric method, one can never expect a uniform symmetry energy coefficients
except the nuclear have the same neutron and proton density difference which result in the
isospin phenomena in HICs. The nuclear density evolves in neutron-rich fragments, thus the
isospin effects illustrates. It is discussed to differ the core and surface regions of a
neutron-rich nucleus, in which the neutron and proton densities differ evenly while in the
surface the neutron and proton densities differ largely \cite{MaCW13isoSB,MaCW11CPC,MaCW09PRC}.
For the neutron-rich nucleus or fragments, we should also expect the evolution of symmetry
energy due to the change of proton and neutron density differences. Based on the equilibrium
assumption, the symmetry energy of identical source is supposed to be same, and $a_{sym}/T$ of
the source obtained should independent on $A$ and $I$. It is revealed that the chemical potential
difference between neutrons and protons, which is an important index of symmetry energy (as in
isoscaling), is also found to vary little in the central collisions and supports the assumption
that the symmetry energy is identical in similar sources \cite{MaCW13isoSB}. The relative uniform
$a_{sym}/T$ is found in prefragments, but is believed to be modified by the decay process and the
$A$ and $I$ dependence of $a_{sym}/T$ is observed in the final fragments \cite{ChenZQ10-iso-sym,PMar12PRCIsob-sym-isos,MaCW13CPC}.
It is also know for neutron-rich nucleus, the surface-symmetry-energy and volume-symmetry-energy should be included
(the coefficients denoted as $b_s$ and $b_v$, respectively). $b_s$ and $b_v$ can be obtained form
$a_{sym}/A$ \cite{MeiHJPG10,MA12CPL09IYRAsbsbv}; and for finite temperatures neutron-rich fragments,
$b_s$ and $b_v$ of can also be obtained from $a_{sym}/AT$ by assuming the $T^2$ dependence of
the coefficients in the mass formula \cite{Ma2013NST}, which are both found coincident with the
theoretical results.

\section{summary}
\label{summary}
In summary, the RCI effects in the $a_{sym}/T$ of fragments in the IYR methods are investigated.
Four RCI approximations are investigated: (1) the M1--RCI adoting the $a_{c}/T$ determined by
IYR(m); (2) the M2--RCI by fitting the difference between IYRs based on the free energy of the
fragment; (3) the M4--RCI by adopting the theoretical Coulomb energy and an IYR temperature $T=2$
MeV; and (4) neglecting the RCI among the related three isobars. The M1--, M2-- and
M3--RCI are found to have relative small values no larger than 0.4. In particular, the
M2--RCI is the smallest one in the three RCIs due to it includes only very small part of the actual RCI. The effects
of RCI in the $a_{sym}/T$ of fragments are also studied. For fragments in the $^{58}$Ni,
and $^{124, 136}$Xe reactions, the M1-- and M4--$a_{sym}/T$ are found to be the lower and upper
limitations of $a_{sym}/T$. Due to the M2--RCI only includes part of the RCI, it enhances the
value of M2--$a_{sym}/T$, which should be called as the effective $a_{sym}/T$ \cite{Huang-PRC11-freeenergy}.
The M4--$a_{sym}/T$ (omitting the RCI), which overlaps the M2--$a_{sym}/T$, indicates
that the enhancement of $a_{sym}/T$ in M2 is very small, and the $m^2_f$ dependence of RCI in
Eq. (\ref{adfit}) should be very small. The omission of RCI among three isobars is verified to
be reasonable. It can also be concluded that the M1-- and M3--RCI actually underestimate the real
value of $a_{sym}/T$ due to the uncertainty introduced by $a_c/T$ and temperature in them.

The M3--$a_{sym}/T$ of some isobaric chains are compared, which is found to decreases when the
fragment becomes more neutron-rich. Relative consistent value is found in the very neutron-rich
fragments, which indicates that the $a_{sym}/T$ of neutron-rich fragments are similar.

\begin{acknowledgments}
This work is supported by the National Natural Science Foundation of China under grants
No. 10905017, the Program for Science \& Technology Innovation Talents in Universities of
Henan Province (13HASTIT046), and Young Teacher Project in Henan Normal University (HNU),
China.
\end{acknowledgments}

\end{document}